\documentclass[twocolumn,aps,preprintnumbers,showpacs]{revtex4}

\usepackage{epsfig}
\usepackage{amsmath}

\def    \beq    {\begin{equation}} \def \eeq    {\end{equation}}
\def    \bea    {\begin{eqnarray}} \def \eea    {\end{eqnarray}}

\def \nn {\nonumber}

\begin{document}

\preprint{hep-th/yymmnnn}

\title{Absence of  resonant decay for  metastable vacua  in gauge theories of scalar fields}
\author{M. Cardella}
\email{matteo@phys.huji.ac.il}
\author{S. Elitzur}
\email{elitzur@vms.huji.ac.il}
\author{E. Rabinovici}
\email{eliezer@vms.huji.ac.il}
\affiliation{Racah Institute of Physics, The Hebrew University of
Jerusalem, 91904, Israel.}

\begin{abstract}
We prove the impossibility of resonant decay of a metastable vacuum in a theory of a scalar field coupled
to a gauge field. Our result extends to gauge theories of scalar fields a recent no go theorem for resonant tunneling
in a pure scalar field theory.

\end{abstract}

\maketitle

\section{Introduction}

\begin{figure}
\centering
\includegraphics[width=8cm]{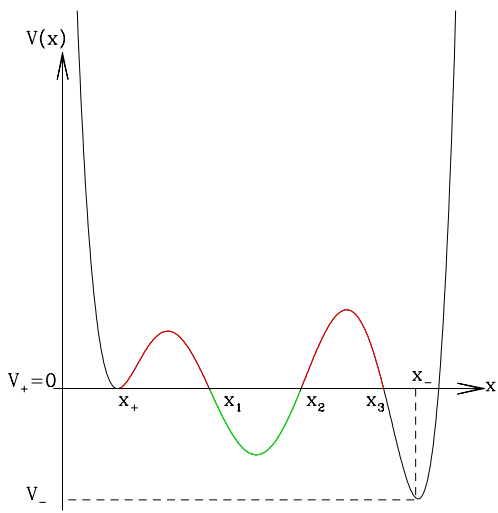}
\caption[]{Double barrier particle decay: In the semiclassical regime a  metastable state with wave function localized
 around $x = x_+$ decays to the true vacuum localized around $x = x_-$.
 If the length of the classically allowed region $ x_1 < x  < x_2$ contains  a half-integer
 number of de Broglie particle wave-length then quantum interference leads to resonant tunneling.}
\label{fig5}
\end{figure}

Understanding in detail  the mobility patterns  in the landscape is a challenging
problem.  In fact rather little is known about it and thus even a study of what may seem
non generic paths is of interest. For example Tye \cite{HenryTye:2006tg},\cite{Tye:2007ja}
has suggested to apply the
phenomena of rapid tunneling \cite{Copeland:2007qf},\cite{Sarangi:2007jb},\cite{Brown:2007ce},\cite{Saffin:2008vi},\cite{Huang:2008jr} to shed a different light on the issue of
the cosmological constant.
Tunneling is generically heavily suppressed.
In quantum mechanics it is known that there  are special barrier configurations for which
the suppression factor is removed and tunneling proceeds as if the barrier were transparent,
this phenomenon is called resonant tunneling.

Although resonant tunneling is a well understood and
observed phenomenon in quantum mechanics \cite{Chang},\cite{Mizuta}, it does not easily
extend  to systems with an infinite number of degrees
of freedom.
 Indeed a no-go theorem for  resonant
tunneling from a metastable vacuum  in a scalar quantum field theory (SQFT) has been recently proved \cite{Copeland:2007qf}.
The aim of the present paper is to study whether
  theories with more structure  then SQFT give a different outcome.
Gauge theories had provided exits out of  no-go theorems
for pure scalar field theories.
One  example is the existence of solitons in $D > 2$ in theories of a scalar field coupled to a gauge
field, which are forbidden for a pure scalar
field theory in $D > 2$ by Derrick theorem \cite{Derrick}.
Motivated  by that, we  will analyze the possibility of resonant tunneling
in theories of scalar fields coupled to gauge fields with several metastable vacua.
We will follow the ideas of the proof  of the no-go theorem for SQFT \cite{Copeland:2007qf},
  generalizing it to a gauge theory in any  space-time dimension.
The result  is that  in a theory
of a scalar field coupled to a gauge field
a homogenous metastable vacuum does not decay in a resonant
fashion.
This provides a no go theorem for resonant vacuum decay
which generalizes \cite{Copeland:2007qf} in allowing the presence
of gauge field with generic coupling  to the scalar field.

\vspace{.2 cm}

The organization of the paper is the following:
In the next section we review resonant tunneling in quantum mechanics,
leaving for the appendix the details of the derivations of the tunneling amplitudes.
We then discuss what it would be required in order to have a similar phenomenon
in quantum field theory, such as the constraints imposed by time-analytic continuation.
In section III we generalize to a generic number of space-time dimensions $D$  the  proof of the no go theorem of \cite{Copeland:2007qf} for resonant
tunneling in SQFT, in order to be able to discuss  resonant  tunneling in
gauge theories of scalar fields.
In the appendix  a derivation of  quantum mechanical tunneling amplitudes and resonance conditions
 for multi-barrier potentials is reviewed.

\section{Resonant tunneling: From QM to QFT}

\begin{figure}
\centering
\includegraphics[width=7cm]{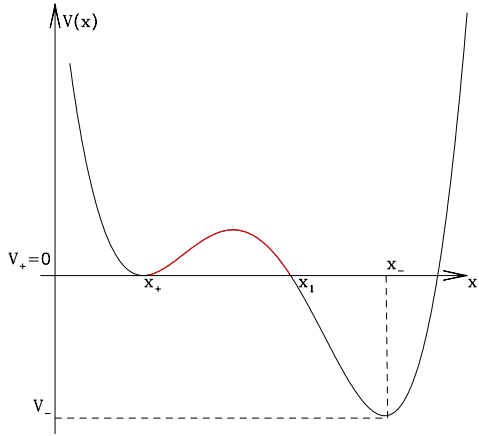}
\caption[]{ In the semiclassical regime, a particle with wave function localized around  a false vacuum $ x = x_+$
can decay to the true  vacuum localized around $ x = x_-$ by
quantum tunneling the barrier $x_+ < x  < x_1 $.}
\label{fig1}
\end{figure}

Quantum interference is typical in systems with a
degenerate set of classical trajectories.
In  the double barrier potential $V(x)$ in figure \ref{fig5}, in the semiclassical regime
a point particle with wave function localized
around the false vacuum  $x = x_+$ will decay to the true vacuum
localized around $x = x_-$. The tunneling process involves an infinite
number of decay paths,
labeled by the  number  of oscillations
in the classical allowed region $x_1< x <x_2$.
Different oscillating  paths can
constructively  interfere
and, for specific particle wave lengths the double potential barrier
becomes   completely transparent.
This phenomenon is called   resonant tunneling,
and it has been observed in various experiments \cite{Chang},\cite{Mizuta}.

 While the  amplitude to decay by tunneling through the single barrier in fig. \ref{fig1}
is at leading order in $\hbar$  exponentially suppressed
by  a   instanton action
\beq
T = \frac{1}{\cosh(S_{I})} \sim   \exp\left(-\frac{1}{\hbar}\int_{x_+}^{x_1} dx \sqrt{2V(x)}\right), \label{1barrier}
\eeq
in the double barrier case, if all the oscillating paths
in  region $ x_{1}< x <x_2$ have the same phase at $x = x_2$,
they create a constructive interference
for the wave function entering
 the second  forbidden region  $ x_{2}< x <x_3$.
Constructive interference
happens  when the distance between the two inversion points
$x=x_1$ and $x=x_2$ contains    an half-integral number of  de Broglie
wave lengths

 \beq
S_{II} = \frac{1}{\hbar}\int_{x_1}^{x_2} dx \sqrt{- 2V(x)} = \left( N + \frac{1}{2} \right) \pi, \label{risona}
 \eeq
 for integers $N$.
Equation (\ref{risona}) denotes a resonance condition, since in this case
 the  amplitude to decay to $x = x_-$  reaches its maximum
modulus
 \beq
|T_{+  -}| = \frac{1}{\cosh(S_{I} - S_{III})}. \label{ris}
\eeq
  $S_{I}$  and  $S_{III}$ are  the instanton actions that dominate
  the amplitude in  the two forbidden regions $x_{+}< x <x_1  $,
and  $ x_{2}< x < x_3 $,
\bea
S_{I} &=&   \frac{1}{\hbar}\int_{x_+}^{x_1} dx \sqrt{ 2V(x)},  \nn \\
S_{III} &=&   \frac{1}{\hbar}\int_{x_2}^{x_3} dx \sqrt{ 2V(x)}.
\eea

 If $S_I \sim S_{III} $ the decay amplitude
 (\ref{ris})  gets close to one  $|T_{+-}|\lesssim 1$, and thus the
 metastable state $x=x_+$ becomes very short living.
For this to happen it is required that $S_I - S_{III} \sim 0$, and that $S_{II}$ satisfies  eq.(\ref{risona}).
 In the semiclassical regime  both  $S_{I}$
and $S_{III}$ are  large numbers in $\hbar$ units,
thus  the  condition   $S_I - S_{III} \sim 0$   requires a large amount
 of fine tuning for the shape of the potential $V(x)$.

\vspace{.2 cm}

 \begin{figure}
\centering
\includegraphics[width=8cm]{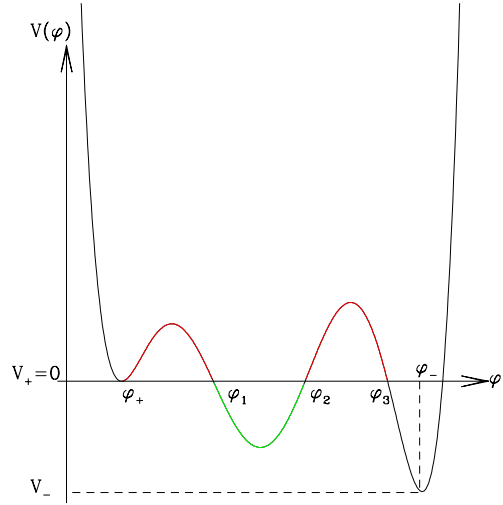}
\caption[]{ Double barrier decay in QFT.
The field in a false vacuum $\varphi = \varphi_{+}$ can decay trough quantum tunneling
toward lower energy density local minima of the effective potential.}
\label{phifig5} \end{figure}

\vspace{.2 cm}

  In  quantum field theory the post-tunneling dynamics would begin  in a localized space-time region,
in a way similar to the one-barrier tunneling field decay discussed in
\cite{Coleman:1977py},\cite{Callan:1977pt}.
The field tunnels in a finite
space-time region by making a quantum jump from its original
value  $\varphi = \varphi_+$
to the final vacuum value  $ \varphi = \varphi_-$, ( see figure \ref{phifig5}).
In the thin wall  approximation, new and old vacua are  separated   by a domain wall,
whose positive energy balances exactly the decrease  in energy
in the limited new vacuum region.
The domain wall will finally  expand by
 converting potential energy
into wall kinetics energy.

\vspace{.2 cm}

The object that computes the  amplitude   per unit of space-time volume
for the field to tunnel through a classically forbidden region   is the Euclidean path integral
\beq
I = \int \mathcal{D}\varphi e^{- S_{E}[\varphi]} \label{EPI}.
\eeq

In the semiclassical approximation the dominant contributions to $I$
 are Euclidean classical fields (instantons).
These Euclidean fields satisfy a global constraint which is the Euclidean version
of energy conservation
\bea
&&\frac{\partial}{\partial t_E} \int d\vec{x} \left [\frac{1}{2}\left(\frac{\partial \varphi}{\partial t_E }  \right)^{2} - \sum_{i=1}^{d} \frac{1}{2}\left(\frac{\partial \varphi}{\partial x_i }  \right)^{2} - V(\varphi) \right]
\nn \\ &=&
\frac{\partial}{\partial t_E}\int d\vec{x} \left[ \frac{1}{2}\left(\frac{\partial \varphi}{\partial t_E }  \right)^{2} - U[\varphi] \right] \nn \\ &=& 0. \label{Ee}
\eea

In  quantum field theory (QFT) in order to have quantum interference
 a periodic field path $\varphi$,  solution of the equations of motion with Lorentzian signature,
would be required.
Resonant tunneling would then follow if its  action $S[\varphi]$
 satisfied
 \beq
 S[\varphi] = \left( N + \frac{1}{2} \right)\pi,\label{quant}
 \eeq
for integer $N$.

This periodic  path  should be connected  via time analytic continuation to
Euclidean paths,  the latter  describing tunneling  in classically
  forbidden regions of the potential.
The question is whether in  quantum field theory
 there are oscillating phenomena that parallel
  the oscillating particle paths in the  region $x_1 < x < x_2 $ of figure \ref{fig5}.

In the classically allowed regions, where the potential for  constant field configuration
is less then the metastable vacuum energy density,
the generalized potential $U[\varphi]$  of some classical field solution could become negative,
(for example this could happen for the potential in fig. \ref{phifig5}
 if  on a sufficiently large space region $\varphi$  assumes a value
 in the interval  $\varphi_1  < \varphi < \varphi_2$).
If this happens one is forced
 to analytic continue the solution to real time $t_E \rightarrow it_E = t$ at the point
 $\tilde{t}_E$  where $\partial_{t_E}\varphi(\tilde{t}_E , \vec{x})=0$.
Analytic continuation in  $\tilde{t}_E = \tilde{t}$   requires
 $\partial_{t_E}\varphi(\tilde{t}_E , \vec{x})=0= \partial_{t}\varphi(\tilde{t} , \vec{x})$.

In the following  we will assume analytic continuation
   of an instanton $\varphi(t_E , \vec{x})$  to Lorentzian time and
 then back to Euclidean time
 \beq
  U[\varphi](t^{1}_{E}) = U[\varphi ](t^{2}_{E}) = 0,
    \ \ \ \ \ \    U[\varphi](t_E) < 0,  \ \ \  t_{E}^{1}<  t_E < t_{E}^2. \nn
 \eeq

  The Lorentzian continuation  $\varphi(t , \vec{x})$  is defined on the
 interval $[t_1 , t_2 ]$, where $t_1 = t_{E}^{1}$ and $t_2 = t_{E}^{2}$.

\vspace{.1 cm}

Together with this  solution there are equally dominant contributions from paths in which   the Euclidean instanton  is continued to
a Lorentzian classical field   $\varphi_{n}(t , \vec{x})$, which is a periodic extension of $\varphi$
for a time period    $(2n + 1)(t_2 - t_1 )$, for integer  $n$.
$\varphi_{n}(t, \vec{x})$ would contribute to the path integral
with action $S[\varphi_{n}]  = S[\varphi]_{(2n + 1)(t_1 , t_2)}  = (2n + 1)S[\varphi]_{(t_1 , t_2)}$, where
\bea S[\varphi]_{ (t_1 , t_2) } = \nn \\
\int_{t_1}^{t_2}dt \int d\vec{x} \left[ \frac{1}{2}\left(\frac{\partial \varphi}{\partial t }  \right)^{2} - \sum_{i=1}^{d} \frac{1}{2}\left(\frac{\partial \varphi}{\partial x_i }  \right)^{2} - V(\varphi)\right],\label{acti}
\eea
thus  creating
quantum interference in the path-integral.

A necessary condition  for
quantum interference in a multi-vacua potential to affect  field decay
is therefore  the existence  of a finite action   classical Lorentzian solution $\varphi$
defined on a   time interval $[ t_1 , t_2 ]$. $\varphi$ has vanishing time derivative
$\partial_{t}\varphi(t_1 , \vec{x}) = \partial_{t}\varphi(t_2 , \vec{x})  = 0 $
in order to be analytically continued to instanton field solution.
In the following, we will check
whether $\varphi$ has non-vanishing action. We consider
resonant decay in  scalar fields theories, already discussed in \cite{Copeland:2007qf},
and  the same issue in gauge theories of scalar fields, which is the
focus of this paper.

\section{Vacuum decay in a landscape potential and the question of quantum interference}

As discussed at the end of the previous section,
 in order to have quantum interference
in vacuum decay  the existence of a
  Lorentzian solution  $\varphi(t, \vec{x})$ defined on a time interval $[t_1 , t_2 ]$ with non-vanishing action satisfying eq. (\ref{quant}) is required.
 Analytic continuation of $\varphi$ to Euclidean signature
  in $t = t_1 = t_2 $  requires
 $\partial_{t}\varphi(t_1 , \vec{x}) = \partial_{t}\varphi(t_2 , \vec{x}) = 0$.
  Moreover
$\varphi$ must respect the total energy constraint.
Since we are studying the decay of the false vacuum, $\varphi$  has to approach at spacial infinity
the original metastable vacuum $\varphi_+$
\beq
\lim_{|\vec{x}|\rightarrow \infty}\varphi(t,\vec{x}) = \varphi_{+}, \label{bc}
\eeq
and therefore
\beq
\lim_{|\vec{x}|\rightarrow \infty}\partial_{\mu}\varphi(t,\vec{x}) = 0.
\eeq

\vspace{.2 cm}

\subsection{No resonant tunneling in  a  scalar field theory}

 Let us consider a scalar field theory in $D = 1 + d$ space-time dimensions
with Lagrangian
\beq
\mathcal{L} = \frac{1}{2} \partial_{\mu} \varphi \partial^{\mu} \varphi    - V(\varphi),
\eeq
where $V(\varphi)$ is a generic potential with various metastable vacua.

  A necessary condition for the field $\varphi$ to have non-vanishing action
     in the interval $[t_1,t_2]$ is the strict
positivity of the following functional

\beq
\mathcal{O}_{L}[\varphi] = \int_{t_1}^{t_2}dx_0 \int d \vec{x} \left[(\partial_{0}\varphi)^2
 + \frac{1}{d}\sum_{i = 1}^{d} (\partial_{i}\varphi)^2 \right]. \label{oscill}
\eeq
In fact, if $\mathcal{O}_{L}[\varphi]=0 $ , being an integral of  a sum of positive terms, the derivatives of the field vanish identically and $S[\varphi] = S[\varphi_+ ]=0$.

By using
 \beq
 \partial_{\mu}\varphi \partial_{\nu} \varphi  =  T_{\mu \nu} + \eta_{\mu \nu}\mathcal{L}
 \eeq
$\mathcal{O}_{L}[\varphi]$ can be written in terms of the  following combinations
  of diagonal  components of the energy-momentum tensor $T_{\mu \nu}$
 \beq
 \mathcal{O}_{L}[\varphi] = \int_{t_1}^{t_2}dx_0 \int d \vec{x} \left[   T_{00} +  \frac{1}{d}\sum_{i = 1}^{d} T_{ii}\right]. \label{o}
 \eeq

 Energy-momentum conservation $\partial^{\mu}T_{\mu i} = 0$ for the $i$-spacial component gives
 \bea
   \partial_{i}T_{ii} &=& \sum_{j \ne i} \partial_{j}T_{ji} - \partial_{0}T_{0i} \nn \\
&=& \sum_{j \ne i} \partial_{j}(\partial_{j}\varphi \partial_{i} \varphi) - \partial_{0} (\partial_{0}\varphi \partial_{i} \varphi). \nn \\
\eea

This implies that
\beq
\partial_{i}\int_{t_1}^{t_2}dx_0 \int \prod_{j \ne i}dx_{j} T_{ii} = \partial_{i}I_{ii} =   0,
\eeq
the $T_{ii}$ density integrated over the space-time domain orthogonal to the $x_i$-direction
and containing  the time interval $(t_1, t_2)$ is constant along $x_i$.
Therefore  $I_{ii}$  can be computed at $x_{i} = \infty$,
where it depends only on $V(\varphi_+)$, since all the field derivatives
 go to zero at infinity
\bea
I_{ii} &=& \lim_{x_{i} \rightarrow \infty }
 \int_{t_1}^{t_2}dx_0 \int \prod_{j \ne i}dx_{j} \cdot \nn \\
&& \left( \frac{1}{2}(\partial_{0}\varphi)^2 +\frac{1}{2}(\partial_{i}\varphi)^2 - \frac{1}{2} \sum_{i = 1}^{d} (\partial_{i}\varphi)^2  - V(\varphi) \right)   \nn \\
 &=&  - \int_{t_1}^{t_2}dx_0 \int \prod_{j \ne i} dx_{j} V(\varphi_+ ). \label{i}
\eea

 On the other hand,  $\partial^{\mu}T_{\mu 0} = 0$ gives
\beq
\partial_{0}T_{00} = \sum_{i = 1}^{d} \partial_{i}T_{i0} = \sum_{i = 1}^{d} \partial_{i}
(\partial_{i}\varphi \partial_{0} \varphi),
\eeq
that integrated all over the space gives energy conservation
\beq
\partial_{0}I_{00} =  \partial_{0} \int d \vec{x} \  T_{00} = 0.
\eeq
The energy equals the energy of our initial state at $t_1$. This being the final configuration of the Euclidean tunneling, its Euclidean energy is the same as the Euclidean energy of the original decaying state. Since both configurations have zero time derivatives, there is no distinction between Euclidean and Lorentzian energy. Assuming a homogeneous original false vacuum  $\varphi = \varphi_{+}$,
\beq
I_{00} = \int  d \vec{x} \  V(\varphi_ + ). \label{0}
\eeq
By using  eq.(\ref{i}) and eq.(\ref{0})
 in the field functional $\mathcal{O}_{L}[\varphi]$  (\ref{oscill}) one obtains

\bea
\mathcal{O}_{L}[\varphi] &=& \int_{t_1 }^{t_2 }d x_0 \  I_{00} + \frac{1}{d }\sum_{i=1}^{d}\int dx_{i} \ I_{ii} \nn \\
&=& \int_{t_1 }^{t_2 }d x_0 \int d \vec{x} \left[ V(\varphi_+) -   \frac{1}{d}\sum_{i=1}^{d} V(\varphi_+ )    \right] \nn \\
&=& 0.\label{canc}
\eea
which shows that  $\partial_{0}\varphi = |\vec{\nabla} \varphi| = 0 $ in the time interval $[t_1 , t_2]$,
thus implying  that the field $\varphi$ has zero action  in the time interval $[t_1 , t_2 ]$.
Moreover, since the field is constant in $[t_1 , t_2]$ it must be equal to  the false vacuum
$\varphi = \varphi_+$.
This shows that an intermediate analytic continuation to Lorentzian signature
is impossible, and the full tunneling event is described by instanton fields in Euclidean signature.
Thus one concludes that for the decaying of a metastable vacuum in a  pure scalar field theory with  a multi-vacua
potential there is no   quantum interference in the
semiclassical approximation. This proves the no go theorem for resonant tunneling in a pure scalar
field theory in a  generic number of space-time dimensions $D$.
Generalization of this result to an arbitrary number of interacting  scalar fields
is straightforward,  since also in this case one can compute each of the  terms in the
 space-like diagonal components of  the energy-momentum
tensor  appearing in (\ref{o}) by going at spacial infinity, where they all equals
minus  the energy of the metastable vacuum.

Recently, \cite{Saffin:2008vi} have shown that special non-homogenous field configurations with finite energy with
respect to a metastable vacuum do exhibit resonant decay. These states circumvent the no-go theorem
of \cite{Copeland:2007qf} in allowing a non-vanishing r.h.s. for  eq.(\ref{0}), which makes  (\ref{canc})
strictly positive, thus allowing for  oscillating in time field solutions.

It is also  interesting  to observe that instantons
 describing motion in forbidden regions are non constant fields connected by analytic continuation
to two real time solutions, the field before and after tunneling, at point
with  vanishing Euclidean time derivative.
Instantons are therefore the Euclidean version of the oscillating
 field forbidden by  the no-go theorem.
The no-go theorem must fail in Euclidean signature, and it does.
 The Euclidean version of  eq.(\ref{oscill}) is
\beq
\mathcal{O}_{E}[\varphi] = \int_{t_E^1 }^{t_E^2 }dx_0 \int d \vec{x} \left[(\partial_{0}\varphi)^2
 + \sum_{i = 1}^{d} (\partial_{i}\varphi)^2 \right]. \label{Eoscill}
\eeq
This functional can be written in terms of the Euclidean stress tensor $T^{E}_{\mu \nu}$ by
summing over the diagonal components of

 \beq
 \partial_{\mu}\varphi \partial_{\nu} \varphi =  T^{E}_{\mu \nu}  + \delta_{\mu \nu}\mathcal{L}_{E}, \label{stress}
 \eeq
where the Euclidean Lagrangian is
\beq
\mathcal{L}_{E} = \frac{1}{2} \partial_{\mu} \varphi \partial_{\mu} \varphi    + V(\varphi).
\eeq
By using  eq.(\ref{stress}) and the Euclidean conservation law $\partial_{\mu}T^{E}_{\mu \nu} = 0$,
one can rewrite (\ref{Eoscill}) as
\beq
\mathcal{O}_{E}[\varphi] = - 2 \frac{D}{D - 2}\int_{t_{E}^1}^{t_{E}^2}dt_E \int d\vec{x} \  V(\varphi(t_E , \vec{x})).
\eeq
The above equation makes sense only when the integral in the  r.h.s. is non-negative.
This is achieved when there is a large enough region of Euclidean space where the field acquires  values
 in classically allowed regions  where $V(\varphi ) < 0$, such
as   the regions  $\varphi_1  < \varphi < \varphi_2 $ and  $\varphi > \varphi_3$
for the potential  in figure (\ref{phifig5}).

\subsection{No quantum interference in the decay of a scalar field coupled to a gauge field}\label{GS}

We study now whether  a theory of a scalar field coupled to  a gauge field
admits  non-constant classical field solutions of Lorentzian time signature that can be
analiticaly continued to Euclidean signature in two separate occasions.

 We consider the Lagrangian
\beq
\mathcal{L} = -\frac{1}{4}F^{a}_{\mu\nu}F^{a \mu\nu} + \frac{1}{2}D_{\mu}\varphi (D^{\mu}\varphi)^{\dag} - V(\varphi).
\label{lag} \eeq

In the semiclassical approximation the Euclidean path integral
   is dominated by  Euclidean classical fields.
These fields satisfy the global constraint

 \bea
 && \partial_{0} \int d \vec{x} \Bigg[ \frac{1}{2}\sum_{i=1}^{d}(F_{0 i}^{a})^2
 + \frac{1}{2}D_{0}\varphi (D_{0}\varphi)^{\dag} \nn \\
  &-& \frac{1}{4}\sum_{i,j=1}^{d}(F_{ij}^{a})^2
 - \sum_{i=1}^{d}D_{i}\varphi ( D_{i}\varphi )^{\dag}  - V(\varphi ) \Bigg] = 0, \nn
\eea
which is an Euclidean version of energy conservation.
We choose   the original metastable state energy equal to zero $V(\varphi_+) = 0$
so that the previous equation gives
\bea
 &&  \int d \vec{x} \Bigg[ \frac{1}{2}\sum_{i=1}^{d}(F_{0 i}^{a})^2
 + \frac{1}{2}D_{0}\varphi (D_{0}\varphi)^{\dag} \Bigg] \nn \\
  &=& \int d \vec{x}  \Bigg[  \frac{1}{4}\sum_{i,j=1}^{d}(F_{ij}^{a})^2
 + \sum_{i=1}^{d}D_{i}\varphi ( D_{i}\varphi )^{\dag}  + V(\varphi ) \Bigg] \nn \\
  &=&: U[\varphi, A_{\mu}^{a} ]. \label{Eeng}
\eea
Given the Euclidean fields $\varphi(t_E , \vec{x})$, $A_{\mu}^{a}(t_E , \vec{x}) $,
  (\ref{Eeng}) is an equation for  $t_{E}$.
For values of the Euclidean fields in regions where $V(\varphi) < 0$,
the r.h.s of eq.(\ref{Eeng}) might become negative $U[\varphi , A_{\mu}^{a} ] < 0$.
We assume  that this is the case on an interval $t_{E}^{1} < t_E  < t_{E}^{2}$,
and check the possibility of a  analytic continuation of the  fields $\varphi(t_E , \vec{x})$ and $A_{\mu}^{a}(t_E , \vec{x}) $
 to Lorentzian signature
$t_E \rightarrow it_{E} = t$ at $t_E^{1} = t_1$ and back to Euclidean signature
$t  \rightarrow -it  = t_{E}$ at  $t_E^{2} = t_2$.
The  Lorentzian classical solutions $\varphi(t,\vec{x})$ and $A_{\mu}^{a}(t,\vec{x})$
are then defined on the interval $[t_1 , t_2 ]$, they dominate the path integral and
 have both vanishing time  derivative at $t=t_1$ and $t=t_2$, as required by
analytic continuation.
In order to check if there is a degeneracy of Lorentzian motions that can
produce interference in the path integral we will check
 if $\varphi(t,\vec{x})$ and $A_{\mu}^{a}(t,\vec{x})$ have a
 non vanishing action in the interval $[t_1 , t_2]$.

To this aim we again use the energy-momentum tensor

\beq
T_{\mu \nu} = F_{\mu }^{\rho a} F_{\rho \nu}^{a} +
 D_{\mu}\varphi (D_{\nu} \varphi)^{\dag} - \eta_{\mu \nu}\mathcal{L},
\eeq

and consider  the following sum over the diagonal components of $T_{\mu \nu}$

\bea
&& \mathcal{O}_{L}[\varphi , A^{a}_{\mu}] = \nn \\
 &&\int_{t_1}^{t_2}dx_0 \int d\vec{x} \left [ D_{0}\varphi (D_{0}\varphi)^{\dag} + \frac{1}{d}\sum_{i = 1}^{d}
D_{i}\varphi (D_{i}\varphi)^{\dag} \right] \nn \\
 &+&  \int_{t_1}^{t_2}dx_0 \int  d\vec{x}  \left[ \frac{d - 1}{d} \sum_{i = 1}^{d}(F^{a}_{0 i})^2
+ \frac{1}{d}\sum_{i,j  = 1}^{d}(F_{i j }^{ a})^2  \right] \nn \\
&=&\int_{t_1}^{t_2}dx_0 \int d\vec{x} \left[ T_{00} +  \frac{1}{d}\sum_{i = 1}^{d}T_{ii}  \right].\label{osgauge}
\eea

 $\mathcal{O}_{L}[\varphi , A^{a}_{\mu}] \ge 0$ is a semi-definite positive functional,
 and a field solution  $(\tilde{\varphi} , \tilde{A}^{a}_{\mu})$ with a  non-vanishing action gives
 $\mathcal{O}_{L}[\tilde{\varphi} , \tilde{A}^{a}_{\mu}] > 0$.
We use  the  component equations of  $\partial^{\mu}T_{\mu \nu} = 0$
 integrated over space-time subregions, in order to show
 that indeed for every classical solution  $\mathcal{O}_{L}[\varphi , A^{a}_{\mu}] = 0$.
This implies that every solution defined on a finite time interval $[t_1 , t_2 ]$
which can be analiticaly continued to Euclidean signature has to be a constant field configuration equal to the false vacuum.
The conclusion will be that field paths describing the full tunneling
 event  are purely Euclidean and therefore there  is no  resonant decay.

\vspace{.3 cm}

The overall $T_{00}$ space integral is set to be equal to the total false vacuum energy $V(\varphi_+)=0$
that is our initial condition in time, therefore the space integral of $T_{00}$ vanishes in the r.h.s of (\ref{osgauge}).

On the other hand,  the component equation $\partial^{\mu}T_{\mu i} = 0$ gives

\bea
\partial_{i}T_{ii} &=& \partial_{0}T_{0i} - \sum_{j \ne i} \partial_{j}T_{ji} \nn \\
&=& \partial_{0} (F_{0}^{\rho a}F^{a}_{\rho i} + D_{0}\varphi (D_{i}\varphi )^{\dag} ) \nn \\
&-& \sum_{j \ne i} \partial_{j} (F_{i}^{\rho a}F^{a}_{\rho j} + D_{i}\varphi(D_{j}\varphi)^{\dag} ).
\eea

By integrating the previous equation over  a space-time  volume orthogonal to the $x_i$ direction
one gets
\bea
&&\partial_{i}\int_{t_1}^{t_2}dx_0 \int \prod_{j \ne i}dx_{j}T_{ii} = \nn \\
&&\int \prod_{j \ne i}dx_{j} \int_{t_1}^{t_2}dx_0   \partial_{0} \left(- \sum_{j \ne i} F_{0 j}^{a}F^{a}_{ji}
 + D_{0}\varphi (D_{i}\varphi )^{\dag} \right) \nn \\
&-& \int_{t_1}^{t_2}dx_0 \int \prod_{j \ne i}dx_{j} \sum_{j \ne i} \partial_{j} (F_{i}^{\rho a}F^{a}_{\rho j} + D_{i}\varphi(D_{j}\varphi)^{\dag} ) \nn \\  &=& 0. \label{iindep}
\eea

 The previous equation states  that the above integral
  is constant along  $x_i$, therefore it
 is  equal to its value for $x_{i} \rightarrow \infty$, where it easier to see that it vanishes,
since all the field derivatives vanish at spacial infinity,
and the potential goes to the false vacuum $V(\varphi_+)=0$.

We have shown that the  r.h.s of eq. (\ref{osgauge}) vanishes, from the same
equation one can read that the two Lorentzian continuations $\varphi(t,\vec{x})$, $A_{\mu}^{a}(t, \vec{x})$
 have a  vanishing
action (\ref{lag})  in the interval $[t_1 , t_2 ]$.
We conclude that there is no quantum interference
  in vacuum decay in a theory  of  a scalar field coupled to a
gauge field. Therefore even in the presence of a gauge field
 a metastable vacuum cannot decay in a resonant fashion.
Generalization of the above result  to a gauge theory of an arbitrary
number of  interacting scalar fields is straightforward, since
all the steps of  the proof we gave  go through also in this case.

\section{Conclusions}
We studied whether in a theory of a scalar field coupled to a gauge field
a homogeneous metastable vacuum can decay in a resonant fashion.
The answer turns out to be  negative and thus our result
generalizes the no go theorem for resonant tunneling in a pure scalar field theory \cite{Copeland:2007qf}.
The proof  follows the line of \cite{Copeland:2007qf},
appropriately  generalized to arbitrary space-time dimensions and to the presence of a gauge field.
 Gauge interaction have exhibited in the past novel non-perturbative phenomena
such as  the existence of solitons in $D > 2$, forbidden
for a pure scalar field theory by Derrick theorem \cite{Derrick}.
 Yet, in the present case they did not come to  rescue,
and  resonant tunneling from a homogeneous false vacuum and its  interesting implications
for the dynamics of the landscape  mentioned in the introduction are also forbidden for
 theories  of  scalar fields coupled to gauge fields.

The authors of the  no go theorem
for resonant tunneling in a pure scalar field theory
 \cite{Copeland:2007qf}
 have recently proposed a way to circumvent
their own no-go theorem \cite{Saffin:2008vi}.
They consider  the decay of
inhomogenous exited states of a metastable vacuum, rather
then the decay of the false vacuum itself.
 These states represent   quite ad hoc
non-homogeneous initial   configurations of the scalar field, and
 whether they can emerge as the outcome of a previous tunneling event in the landscape it remains to be explained.
Besides the need to supply a  natural mechanism that makes the bubbles of  \cite{Saffin:2008vi}
 to contract,  one should also  estimate  the occurrence of such events in the landscape.
 This  would be an essential ingredient
 in order to determine whether
 the proposal of \cite{Saffin:2008vi} represents an
 actual possibility for rapid vacuum decay.  As stressed in \cite{Saffin:2008vi}, the present no go theorem does not apply to non homogeneous initial states. Notice that if indeed non homogeneous initial states can have resonant tunneling, solitonic solutions present in gauge theories may naturally supply such initial states as ground states in given topological sectors.

\section{Appendix: Wave mechanics methods for tunneling amplitudes
through multiple barriers}

\subsection{ Tunneling  through one barrier potential}

\begin{figure}
\centering
\includegraphics[width=8cm]{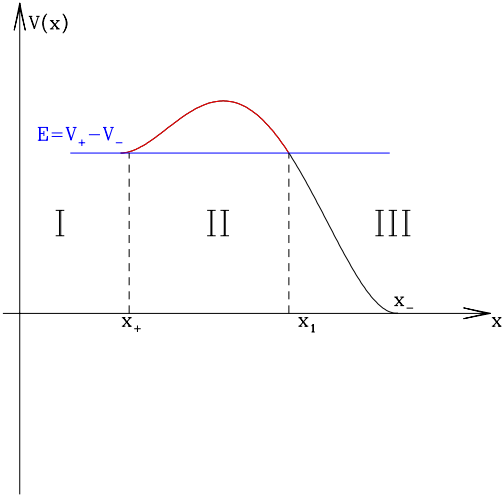}
\caption[]{A particle in region I  with energy  $E = V_+ - V_-$ has a non vanishing probability to tunnel
 to region III   through
the classically forbidden potential  barrier of region II}
\label{fig2} \end{figure}

Let us consider the tunneling problem  represented in figure \ref{fig2},
the semiclassical wave function for a particle of unit mass with energy
$E$ in region I is given by

\beq
\Psi_{I}(x) = \alpha_{I}^{R}\frac{e^{iS_{0}(x)/\hbar}}{ (2(E  -  V(x)))^{1/4}} + \alpha_{I}^{L}\frac{e^{-iS_{0}(x)/\hbar}}{(2(E - V(x)))^{1/4}},\label{waveI}
\eeq

where
\beq
S_{0}(x) = \int_{x_+}^{x}  dx \sqrt{2( E - V(x))}
\eeq
is the reduced action.

In order to compute the tunneling probability
one has to compare the semiclassical wave functions in  region I with
that one in region III, beyond the barrier.
Semiclassical approximation breaks down near the classical turning point
$x=x_+$, and in fact (\ref{waveI}) is divergent there.
The two expressions for the semiclassical wave functions
outside the barrier (region I) and inside the barrier (region II),
can be obtained  by solving the Schroedinger
equation near $x = x_+$, with the linearized potential $2(E - V(x)) \sim 2F_{0}(x - x_+ )$.
Semiclassical approximation is valid on the region
\beq
|x - x_+ | >> \left(\frac{\hbar^{2}}{2F_0}\right)^{1/3},
\eeq
(see for example \cite{Merzbacher}), therefore by comparing the
 two  asymptotic expansion of the solution on the left and
 on the right of the turning point $x=x_+$
 one can obtain a relation between the
 semiclassical wave functions in the two regions,
 called connection formula.

It is amusing to see how through  a simple method one can recover the same connection formulae
without the knowledge of the solution of  the Schroedinger equation for the linearized
potential near the inversion point.
Starting with the semiclassical wave function (\ref{waveI})   in the allowed region I,
one can obtain the correct expression for the semiclassical wave function
beyond the turning point   in region II  by circumventing  $x = x_+$ in the complex plane along a semicircular path
 centered in $x = x_+$:   $x - x_+ \rightarrow e^{i\varphi}( x - x_+ )$.
 There are two ways to circumvent the turning point, a clockwise path $\varphi = -\pi$
 and the anti-clockwise $\varphi = \pi$ one, each way producing a different
 wave function inside the barrier.
 It turns out that the correct   wave function inside the barrier
 is given  by   summing the two contributions:

\bea
\Psi_{II}(x) =  e^{-\Theta} \left(  \alpha_{I}^{R}e^{-i\pi/4} + \alpha_{I}^{L}e^{i\pi/4}\right)              \frac{e^{-S_{0}^{'}(x)/\hbar}}{(2(V(x) - E))^{1/4}} \nn \\
 + e^{\Theta} \left(  \alpha_{I}^{R}e^{i\pi/4} + \alpha_{I}^{L}e^{-i\pi/4}\right)             \frac{e^{S_{0}^{'}(x)/\hbar}}{(2(V(x) - E))^{1/4}}, \nn \\ \label{waveII}
\eea
where
\beq
S_{0}^{'}(x) = \int_{x_1}^{x}  dx \sqrt{2(V(x) - E)},
\eeq

and
 $$\Theta = \frac{1}{\hbar}\int_{x_+}^{x_1} dx \sqrt{2(V(x) - E)}.$$

 By using  $\Psi^{II}$ given by eq. (\ref{waveII}), one can reach region III  by circumventing
on the complex plane  the turning point  $x = x_2$ with two semicircular  paths centered in $x=x_2$.
  The wave function in  region III $x > x_2 $ outside the barrier
 is then given by
\beq
\Psi_{III}(x) = \alpha_{III}^{R}\frac{e^{iS_{0}^{''}(x)/\hbar}}{ (2(E - V(x)))^{1/4}} + \alpha_{III}^{L}\frac{e^{-iS_{0}^{''}(x)/\hbar}}{(2(E - V(x)))^{1/4}},\label{waveIII}
\eeq
with
\beq
S_{0}^{''}(x) =  \frac{1}{\hbar}\int_{x_1}^{x} dx \sqrt{2(E - V(x))}.
\eeq

 The relation between the wave function coefficients in region I and III can be cast
 in the following nice formula

\bea
\begin{pmatrix}
\alpha_{III}^{R}\\
\alpha_{III}^{L}\\
\end{pmatrix}
&=&
\begin{pmatrix}
\cosh\Theta  &  -i \sinh \Theta \\
 i \sinh \Theta &  \cosh \Theta  \\
\end{pmatrix}
\begin{pmatrix}
\alpha_{I}^{R}\\
\alpha_{I}^{L}\\
\end{pmatrix} \nn \\
&=&
\mathcal{R}_{\Theta}\cdot
 \begin{pmatrix}
\alpha_{I}^{R}\\
\alpha_{I}^{L}\\
\end{pmatrix},
\eea

which shows that the wave function which enters from region I
  into the  forbidden region II,  it reappears in region III as "rotated" by the matrix
$\mathcal{R}_{\Theta}$

\beq
\mathcal{R}_{\Theta} =
\begin{pmatrix}
\cosh\Theta  &  -i \sinh \Theta \\
 i \sinh \Theta &  \cosh \Theta  \\
\end{pmatrix}.
\eeq

The factor $i$ in the above  "rotation" matrix ensures that its determinant is one,
a necessary condition to preserve the wave function norm.

The probability for transmission through a single barrier
is  given  by

\beq
|T|^{2} = \left| \frac{\alpha_{III}^{R}}{\alpha_{I}^{R}} \right|^{2} = \frac{1}{(
\cosh\Theta)^2},
\eeq
and in  the semiclassical approximation that we are considering  $\Theta >> 1$

 \bea
|T|^2 &=& \frac{1}{(\cosh \Theta)^2} \sim 4 e^{-2 \Theta } \nn \\
 &=& 4 \exp \left( -\frac{2}{\hbar}\int_{x_1}^{x_2} dx \sqrt{2(V(x) - E )}\right).
\eea
This last formula reproduces  the tunneling probability given in eq. (\ref{1barrier}).

\subsection{Tunneling  through a double barrier potential}

\begin{figure}
\centering
\includegraphics[width=8cm]{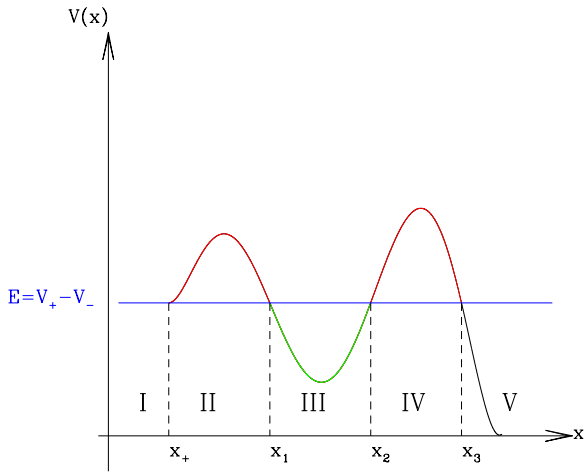}
\caption[]{Double barrier potential: Due to quantum interference there is a finite spectrum of energies
such that the probability to tunnel through the double barrier is higher than
 the probability to tunnel through   a single barrier. This is due to the presence of the intermediate
 classically allowed region III, where the various oscillating paths between the classical turning points
 $x=x_1$, $x=x_2$ can quantum interfere in a constructive way, giving rise to resonant tunneling.}
\label{fig6}
\end{figure}

For a double barrier tunneling (fig. \ref{fig6})
one can compute the full "rotation" matrix which rotates the wave function of region I
into the wave function of region V

\beq
\vec{\alpha}_{out} = \mathcal{R}_{\Theta_1}\mathcal{R}_{\delta}\mathcal{R}_{\Theta_2}\vec{\alpha}_{in},
\eeq
where $\mathcal{R}_{\Theta_1}\mathcal{R}_{\delta}\mathcal{R}_{\Theta_2}$
is given by

\bea
 \begin{pmatrix}
\cosh\Theta_1  &  -i \sinh \Theta_1 \\
 i \sinh \Theta_1 &  \cosh \Theta_1  \\
\end{pmatrix}
\begin{pmatrix}
e^{i\delta}  &  0 \\
 0 &  e^{-i\delta}  \\
\end{pmatrix}
\begin{pmatrix}
\cosh\Theta_2  &  -i \sinh \Theta_2 \\
 i \sinh \Theta_2 &  \cosh \Theta_2  \\
\end{pmatrix} \nn \\
= \cos \delta
\begin{pmatrix}
\cosh( \Theta_1 +\Theta_2 )  &  -i \sinh( \Theta_1 +\Theta_2 ) \\
 i \sinh( \Theta_1  +\Theta_2 )  &  \cosh ( \Theta_1 +\Theta_2 ) \\
\end{pmatrix} \nn \\
+ \ i\sin \delta
\begin{pmatrix}
\cosh( \Theta_1 -\Theta_2 )  &  -i \sinh( \Theta_1 -\Theta_2 ) \\
 i \sinh( \Theta_1 -\Theta_2 )  &  \cosh ( \Theta_1 -\Theta_2 ) \\
\end{pmatrix}.\nn
\eea

 $$\delta = \frac{1}{\hbar}\int_{x_1}^{x_2} dx \sqrt{2(E -V(x))}$$
is the interference phase responsible for resonant tunneling, while
 $$\Theta_1 = \frac{1}{\hbar}\int_{x_+}^{x_1} dx \sqrt{2(E - V(x))},$$ and
$$\Theta_2 = \frac{1}{\hbar}\int_{x_2}^{x_3} dx \sqrt{2( E -  V(x))},$$
are the corresponding (hyperbolic) angles in the
classically forbidden regions.

The transmission coefficient for double barrier tunneling is obtained
 by  $T = \alpha_{out}^{R} / \alpha_{in}^{R}$, with the condition $\alpha_{out}^{L} = 0$.

Computation  gives the tunneling probability

\beq
|T|^{2} = \frac{1}{\cos^{2} \delta \cosh^{2}(\Theta_1 + \Theta_2 ) +  \sin^{2} \delta \cosh^{2}(\Theta_1 - \Theta_2 )}.
\eeq

 The   term in $\cos \delta$ corresponds to the product of the decay probabilities,
 it is the  non-resonant contribution, dominant  in the  limit $\delta = 0$,
  when interference in the intermediate region plays no role. For $ \delta = 0$   the two  hyperbolic
    angles $\Theta_1$ and $\Theta_2$
sum together giving the sum of the rotations.
The  term in $\sin \delta$  is the effect  of the  quantum interference,
and it depends on the  difference  between $\Theta_1$ and $\Theta_2$.

 The tunneling probability can be closed  to one in the resonant condition
 \beq
  \delta = \left(n + 1/2 \right)\pi, \qquad  \Theta_1 \sim \Theta_2.
 \eeq
Notice that $\Theta_1 \sim \Theta_2$ requires a large amount of fine tuning,
since in the semiclassical regime that we are considering
 both $\Theta_1$  and  $\Theta_2$  are large numbers in $\hbar$ units.

\subsection{Tunneling through  $N$ barriers}

Given $N$ barriers separated by $N-1$ classically allowed regions,
the matrix which connects the wave functions on the  two regions  extending to infinity
is given by  $ \mathcal{R}_N = \prod_{k=1}^{N}\mathcal{R}_{\Theta_{k}}\mathcal{R}_{\delta_{k}}$

\beq
\mathcal{R}_N = \prod_{k=1}^{N}
    \begin{pmatrix}
\cosh\Theta_k  &  -i \sinh \Theta_k \\
 i \sinh \Theta_k &  \cosh \Theta_k  \\
\end{pmatrix}
\begin{pmatrix}
e^{i\delta_K}  &  0 \\
 0 &  e^{-i\delta_K}  \\
\end{pmatrix},
\eeq

where
\beq  \Theta_k = \frac{1}{\hbar}\int dx \sqrt{2V(x)},
\eeq
and
\beq  \delta_k = \frac{1}{\hbar}\int  dx \sqrt{- 2V(x)}.
\eeq

There may be various resonant tunneling conditions, the simplest
one  corresponds to the existence of a energy bound state
common to all the $N-1$ classical allowed regions.
If the incoming particle has this bound state energy
 the total tunneling probability turns out to be

\beq
|T|^{2} = \frac{1}{\cosh^{2}\left(- \sum_{k=1}^{N}(-)^{k}\Theta_{k}\right)},
\eeq

and resonant tunneling occurs for
\beq
\sum_{k=1}^{N}(-)^{k}\Theta_{k} = 0. \label{multires}
\eeq
This resonance condition extends to a generic number of barriers $N$
the resonance  condition
\beq
\Theta_1 - \Theta_2 = 0,
\eeq
 which has been obtained for the double barrier potential in figure \ref{fig6}.

\section{Acknowledgments}
M.C. would like to thank A. Schwimmer for stimulating discussions.
The work of M.C. was supported by Superstring  Marie Curie Training Network under the
contract MRTN-CT-2004-512194.
The work of S.E.  was partially supported by the Israel Science
Foundation, the
Einstein Center in the Hebrew University, and by a grant of DIP (H.52).
The work of E.R. was partially supported by
the European Union Marie Curie RTN network under contract MRTN-CT-2004-512194,
the American-Israel Bi-National Science Foundation, the Israel Science Foundation, The
Einstein Center in the Hebrew University, and by a grant of DIP (H.52).


\begin{thebibliography}{99}




\bibitem{HenryTye:2006tg}
  S.~H.~Tye,
  ``A new view of the cosmic landscape,''
  arXiv:hep-th/0611148.

\bibitem{Tye:2007ja}
  S.~H.~Tye,
  ``A Renormalization Group Approach to the Cosmological Constant Problem,''
  arXiv:0708.4374 [hep-th].

\bibitem{Coleman:1980aw}
  S.~R.~Coleman and F.~De Luccia,
  ``Gravitational Effects On And Of Vacuum Decay,''
  Phys.\ Rev.\  D {\bf 21}, 3305 (1980).

\bibitem{Copeland:2007qf}
  E.~J.~Copeland, A.~Padilla and P.~M.~Saffin,
  ``No resonant tunneling in quantum field theory,''
  arXiv:0709.0261 [hep-th].

\bibitem{Sarangi:2007jb}
  S.~Sarangi, G.~Shiu and B.~Shlaer,
  ``Rapid Tunneling and Percolation in the Landscape,''
  arXiv:0708.4375 [hep-th].

\bibitem{Brown:2007ce}
  A.~R.~Brown, S.~Sarangi, B.~Shlaer and A.~Weltman,
  ``A Wrinkle in Coleman - De Luccia,''
  Phys.\ Rev.\ Lett.\  {\bf 99}, 161601 (2007)
  [arXiv:0706.0485 [hep-th]].

\bibitem{Saffin:2008vi}
  P.~M.~Saffin, A.~Padilla and E.~J.~Copeland,
  ``Decay of an inhomogeneous state via resonant tunnelling,''
  arXiv:0804.3801 [hep-th].

\bibitem{Huang:2008jr}
  Q.~G.~Huang and S.~H.~Tye,
  ``The Cosmological Constant Problem and Inflation in the String Landscape,''
  arXiv:0803.0663 [hep-th].

\bibitem{Chang}
L.Chang, L. Esaki and R. Tsu, Appl. Phys. Lett. {\bf 24} (1974) 593.

\bibitem{Mizuta}
H. Mizuta and T. Tanoue \emph{The Physics and Applications of Resonant Tunneling Diodes},
Cambridge University Press, Cambrbridge (1995).

\bibitem{Derrick}
G.H. Derrick,
J. Math. Phys. 5, 1252 (1964)

\bibitem{Coleman:1977py}
  S.~R.~Coleman,
  ``The Fate Of The False Vacuum. 1. Semiclassical Theory,''
  Phys.\ Rev.\  D {\bf 15}, 2929 (1977)
  [Erratum-ibid.\  D {\bf 16}, 1248 (1977)].




\bibitem{Callan:1977pt}
  C.~G. ~Callan and S.~R.~Coleman,
  ``The Fate Of The False Vacuum. 2. First Quantum Corrections,''
  Phys.\ Rev.\  D {\bf 16}, 1762 (1977).

\bibitem{Merzbacher}
E. Merzbacher, Chapter 7, Quantum Mechanics, 2nd Edition, John Wiley, 1970.




\end{thebibliography}
 \end{document}